\documentclass[twocolumn]{aastex63}

\newcommand{\kms}{\rm{\,km\,s}^{-1}}
\newcommand{\vtan}{V_{\rm{TAN}}}

\received{2019 December}
\revised{2020 April}
\accepted{2020 May}

\shorttitle{Blue stragglers in the halo}
\shortauthors{Casagrande Luca}

\begin{document}

\title{Connecting the local stellar halo and its dark matter density to dwarf
  galaxies via blue stragglers}

\correspondingauthor{Luca Casagrande}
\email{luca.casagrande@anu.edu.au}

\author[0000-0003-2688-7511]{Luca Casagrande}
\affiliation{Research School of Astronomy and Astrophysics, Mount Stromlo Observatory, Australian National University, ACT 2611, Australia}
\affiliation{ARC Centre of Excellence for All Sky Astrophysics in 3 Dimensions (ASTRO3D), Australia} 

\begin{abstract}
  The {\it Gaia} H-R diagram shows the presence of apparently young stars at
  high
  tangential velocities. Using a simple analytical model, I show that these
  stars are likely to be blue stragglers. Once normalized to red giant stars,
  the fraction of nearby halo blue stragglers is of order 20 percent, and 
  remarkably close to that measured in dwarf galaxies. Motivated by this
  similarity, I apply to field blue stragglers scaling relations
  inferred from blue stragglers in dwarf galaxies. Doing this for
  the Milky Way halo returns an average stellar density of
  $(3.4 \pm 0.7) \times 10^{-5} M_{\odot}/\rm{pc}^3$ and a dark matter density
  of $\simeq 0.006^{+0.005}_{-0.003}\,M_{\odot}/\rm{pc}^3 \simeq 0.22^{+0.20}_{-0.10}\,\rm{GeV}/\rm{cm}^3$ within 2 kpc from the Sun.
  These values compare favorably to other determinations available in the
  literature but are based on an independent set of assumptions. 
  A few considerations of this methodology are discussed, most notably
  that  the correlation between the dark matter halo core density and stellar
  mass seen in dwarf galaxies seems to hold also for the nearby Milky Way
  halo.
\end{abstract}


\section{Introduction} \label{sec:intro}

Recent results have shown that the close binary fraction ($P\lesssim
10^{4}$~days and $a\lesssim 10$~AU) of solar-type stars
is anticorrelated with metallicity \citep{mo,elb}, implying that most
solar-type stars with [Fe/H]$<-1$  will interact with a close binary. While the
implications of this results are manyfold, here I focus on the rate of
blue-straggler stars (BSS). First observed by \cite{sandage53} in the globular
cluster M3 as an apparent extension of the classical main sequence, BSS are
now believed to be the product of mass transfer and/or merger in close binaries
or multiple-star systems \citep[e.g.][]{knigge,santana}. The metallicity
anticorrelation of close binaries implies that the fraction of BSS is
expected to increase at decreasing metallicity \citep{wyse20}. This means that
the blue
stragglers population should become more prominent when moving to increasingly
metal-poor and old populations (so that BSS can be readily identified
populating the left-hand side of the turnoff), such as moving from the thin
to the thick disk and
halo. Here I use data from {\it Gaia} DR2 \citep{gaiaDR2} and simple
analytical considerations to show that this identification is indeed possible.
The fraction of BSS found at high tangential velocities is in excellent
agreement with that measured in dwarf spheroidal and ultrafaint dwarf
galaxies. For these stellar systems, the fraction of BSS is proportional to
their total stellar mass \citep{santana}, which then correlates to their dark
matter halo core density. Here I show that by applying these relations to the
BSS identified in the Galactic stellar
halo, it is possible to obtain measurements of the local stellar and dark
matter density that are in agreement with those derived by other means. This
result supports the assumption of applying to the halo of a spiral galaxy like
the Milky Way scaling relations for BSS in dwarf galaxies. Thus, field BSS
might be able to provide a new diagnostic to study the stellar and dark matter
density in the halo. A number of methodologies have been developed over the
years to measure these two densities, which are central, e.g., to guide direct
dark matter detection experiments, to understand the formation of the Milky
Way and to place it in the cosmological context with other, similar-mass
galaxies \citep[e.g.,][]{read14,deason,desalas}. However, these measurements
are far from definitive. For the stellar density in the halo, a wide range of
local normalizations have been reported in the literature. Local dynamical dark
matter density measurements are strongly affected by the imperfect knowledge of
the baryonic contribution, and in spite of the data from Gaia, different
analyses still return dissimilar results \citep[e.g.,][]{s18,desalas}. The use
of blue stragglers to derive stellar and dark matter densities is explored here
by developing a methodology that is based on assumptions largely different
from those used by other methods.

\section{Blue Straggler selection}\label{sec:data}

From {\it Gaia} DR2 I retrieve all stars satisfying conditions (1), (2) and
(3) of \cite{arenou}, and with parallax errors below 10 percent, totaling
$67.7$ million objects. These
requirements are also used in \cite{gaiaHR} to study the fine structure of the
Hertzsprung-Russell diagram, removing most of the artifacts while still
allowing us to see the imprint of genuine binaries. Avoiding selection against
binaries might be relevant given that \cite{ps00} concluded that a
significant fraction of BSS are binaries. For this same reason I do not impose
a threshold on the Renormalized Unit Weight Error (RUWE, technical note
GAIA-C3-TN-LU-LL-124-01), and note that the present sample has a median RUWE
of $1.0$, with 97\% of the stars having RUWE$<1.4$. 

BSS are identified in a fashion similar to \cite{santana} and normalized to
the number of red giant branch (RGB) stars identified in a similar range of
absolute magnitudes. The advantage of using the number count of stars selected
in a similar range of intrinsic luminosities is that their ratio is 
largely insensitive to selection effects stemming from the Malmquist bias
\citep{malmquist}. 

\begin{figure*}
\begin{interactive}{animation}{video.mp4}
\epsscale{1.2}
\plotone{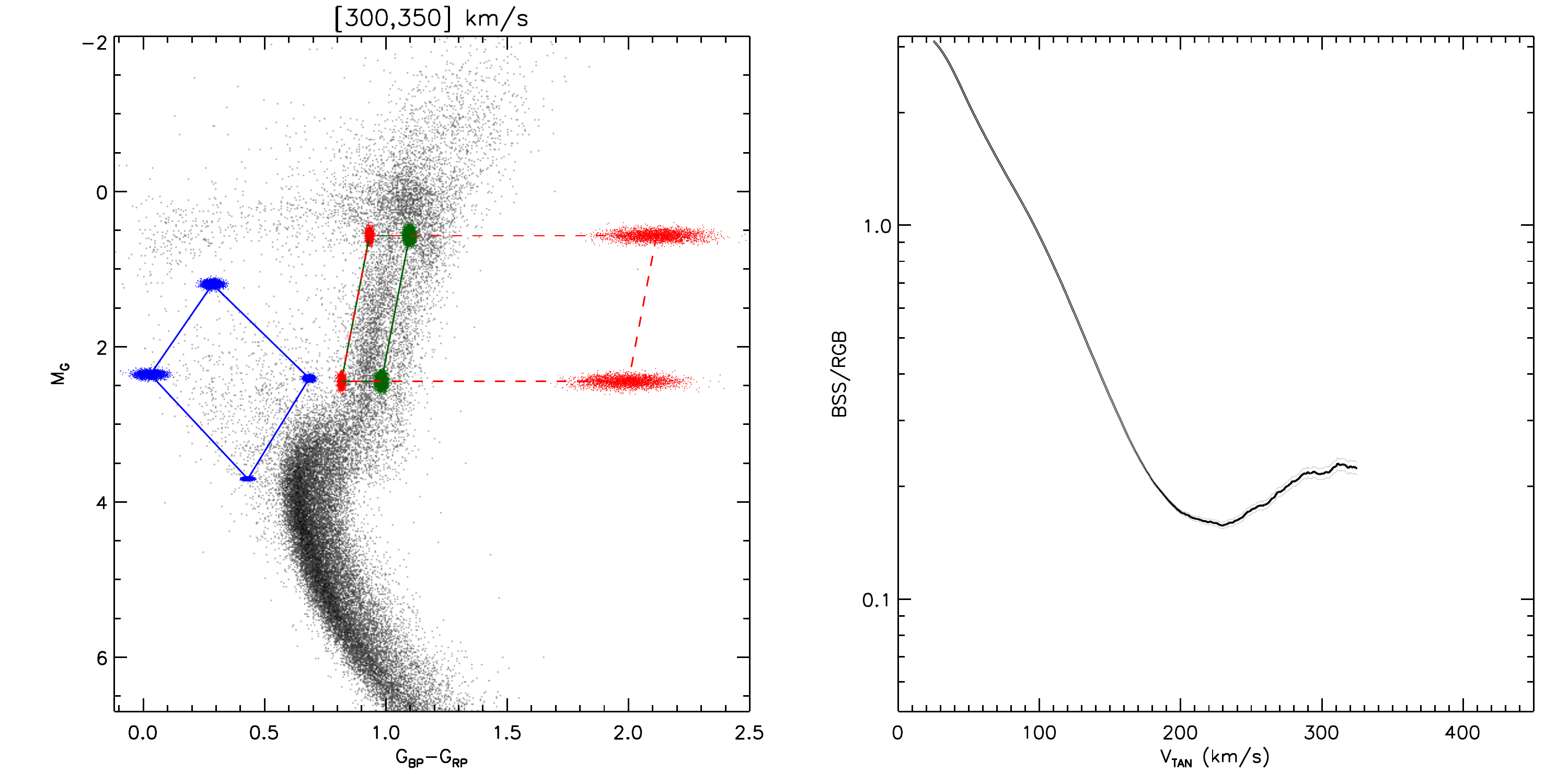}
\end{interactive}
\caption{{\it Left panel:} color-magnitude diagram for the {\it Gaia} sample
  having tangential velocities ($V_{\rm{TAN}}$) in the range indicated on the
  top of the panel.
  Colors and magnitudes have been corrected for reddening as described in the
  text. The blue and red boxes are used to identify BSS and RGB stars,
  respectively. The green box is used to identify bRGB stars (i.e.\,RGB stars
  on the {\it Gaia} blue
  sequence; see description in Section \ref{sec:norma}). Scattered blue, red,
  and green points
  are the range of values explored for the 10,000 Monte Carlo realizations
  described in the paper. 
  {\it Right panel:} black line shows the ratio of stars falling into the BSS
  and RGB boxes when selecting them in a moving $V_{\rm{TAN}}$ boxcar of width
  $50\kms$. Solid gray lines are 1
  sigma Poisson error bars. The point with the highest $V_{\rm{TAN}}$ in this
  figure is obtained doing the ratio of the BSS and RGB stars shown on the
  left-hand panel. 
  An animation of these panels, with a $V_{\rm{TAN}}$ boxcar running from 0 to
  $500\kms$ is available. The duration of the video is 50 seconds.
\label{fig:video}}
\end{figure*}

BSS and RGB stars are defined as stars falling within the blue and red boxes
shown in the color-magnitude diagram of Figure \ref{fig:video}. These boxes are
  obtained following \cite{Gaia_evans} to convert $gr$ magnitudes
  from
  \cite{santana} into the {\it Gaia} photometric system, plus small zero-points
  shifts to optimize these boxes with the actual position occupied by BSS and
  RGB stars in the {\it Gaia} H-R diagram. For the RGB box, the boundary at
  cool temperatures is extended to colors redder than in \cite{santana}, to
  account for a range of metallicities and ages in the Galactic disk that is
  much larger than that encountered in dwarf galaxies or globular clusters.
  This extension is, however, irrelevant when dealing with stars at high
  tangential velocities, which occupy the leftmost position on the RGB box
  (see also animation associated with Figure \ref{fig:video}).
{\it Gaia}
colors and magnitudes of all stars have been corrected for reddening using a
rescaled version of the \cite{sfd} map as described in \cite{kunder}, with
reddening coefficients from \cite{cv18}.

The video associated with Figure \ref{fig:video} shows how the ratio of the
number of stars falling into the BSS and RGB boxes varies when selecting stars
with different tangential velocities. As
$V_{\rm{TAN}}$ increases, the dominant stellar population changes from the thin
to the thick disk, until the ratio remains constant at a value of $0.2-0.3$ for
tangential velocities that are typical of halo stars.

\begin{figure*}
\epsscale{1.2}
\plotone{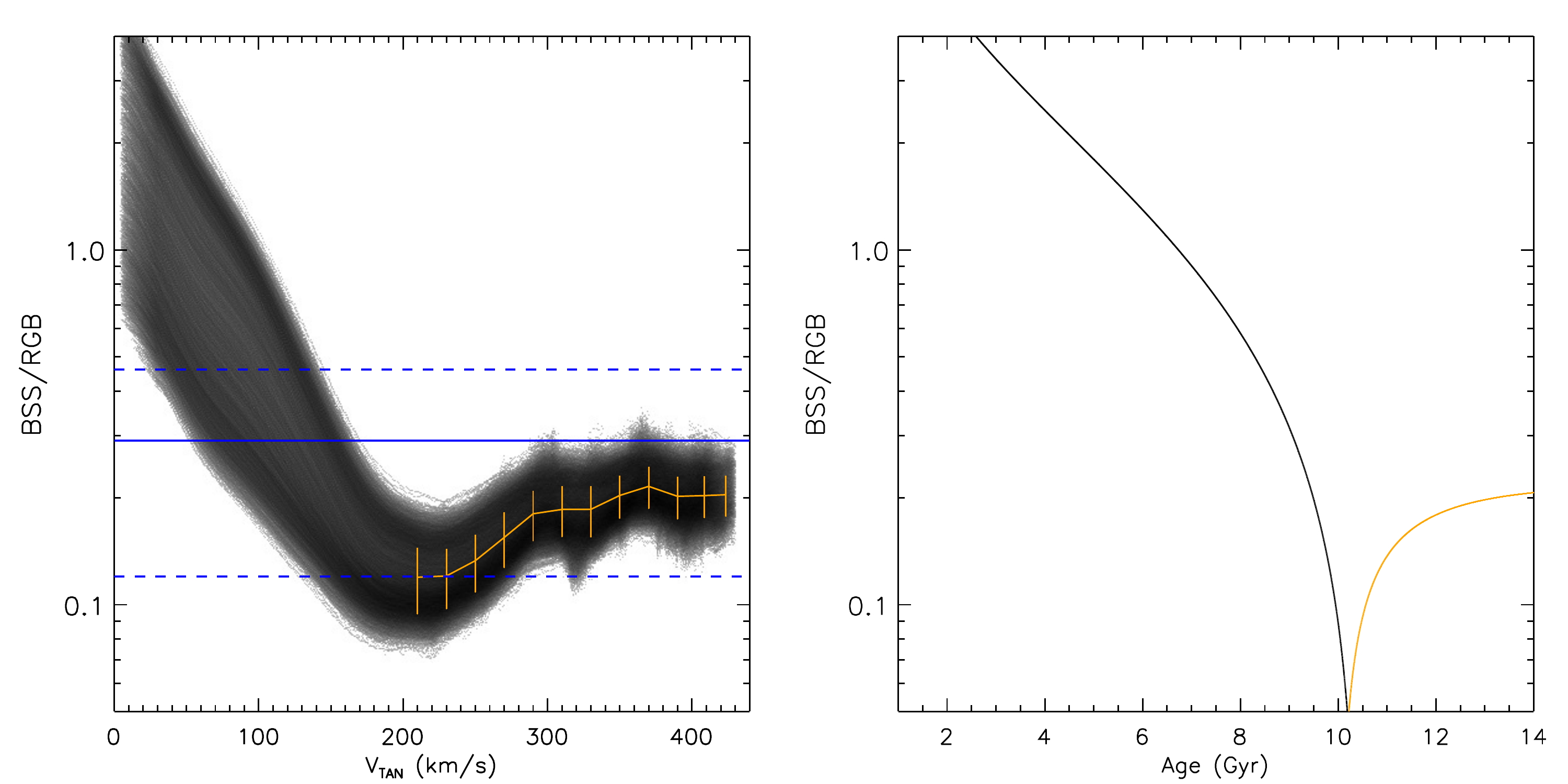}
\caption{{\it Left panel:} BSS-to-RGB ratio as a function of $V_{\rm{TAN}}$
  for 10,000 Monte Carlo realizations varying parameters as described in the
  text (gray lines).
  The orange line is the mean and standard deviation of all realizations above
  $200\kms$. The solid blue line is the BSS-to-RGB ratio measured by
  \cite{santana} for dwarf galaxies, with 1 sigma uncertainties indicated by the
  dashed lines. {\it Right panel:} analytic models of equations \ref{eq:int}
  (black) and \ref{eq:kroupa} (orange).\\
\label{fig:mcanal}}
\end{figure*}

The box used to identify BSS suffers from contamination from main-sequence
stars populating this region of the H-R diagram. This effect is very strong in a
young stellar population like the thin disk \citep[say, $V_{\rm{TAN}}<40\kms$,
  see, e.g.,][]{gaiaHR} and decreases when moving toward older populations like
the thick disk and halo. Figure \ref{fig:video} is obtained using all stars
without any
restriction on their Galactic latitude $b$, or height above the Galactic plane
$Z$.
Another source of uncertainty is due to the arbitrary definition of the BSS and
RGB boxes, as well as to the reddening corrections applied. To account for all
these uncertainties, I run 10,000 Monte Carlo realizations where each time
I randomly changed the boundaries of the BSS and RGB boxes by up to
several hundreds of
mag, reddening by 20 percent, let the width of the boxcar vary anywhere between
$10$ and $60\kms$, and considered only stars with heights above the Galactic
plane varying in the range $0<Z<400$ pc. The results are shown in the left
panel of
Figure \ref{fig:mcanal}. At low $V_{\rm{TAN}}$ the ratio of stars falling into
the BSS and RGB boxes varies quite substantially, and this is largely driven
by the adopted cuts in $Z$. The higher stars are above the Galactic plane, the
lower is the contamination from young thin disk stars that otherwise would
fall into the BSS box. Therefore, the ratio of stars into the BSS and RGB
boxes decreases.
I verified that very similar results are obtained if doing a cut in the
projected height above the plane, i.e. Galactic latitude $b$, instead of $Z$.

Remarkably, the trend reverses for tangential velocities around $200\kms$
  and the ratio becomes nearly constant above $\sim 300\kms$, with very low
scatter independently of the set of parameters of each Monte Carlo realization.
\newpage
\section{Analytical model}

The number of main-sequence stars that contaminate the BSS box relative
to the RGB one can be modeled generating a synthetic stellar population, but
this is beyond the scope of this analysis. Instead, simple analytical
considerations suffice to understand the trend.

The number of main-sequence stars that fall into the BSS box at any given
time is given by
\begin{equation}
N_{\rm{BSS}} = \int_{M_1}^{M_2} \phi(m) \textrm{d}m 
\end{equation}
where $\phi(m)$ is the Initial Mass Function (IMF), and $M_1$ and $M_2$ are
minimum and maximum stellar mass in the BSS box. For stars above roughly a
solar mass, the IMF slope of \cite{salpeter} can be safely used, i.e.
$\phi(m)\propto m^{-2.35}$.
Similarly, the number of stars that
are on the RGB phase at a given time is given by all stars massive enough to
have evolved off the main sequence, i.e. with masses higher than the
time-dependent turnoff mass $M(t)$. The latter can be readily derived from
the relationship between main-sequence lifetime and stellar mass
\cite[e.g.][]{kippe}:
\begin{equation}\label{eq:kippe}
\frac{M(t)}{M_{\odot}}\sim\left( \frac{t}{t_{\odot}} \right)^{-1/3}.
\end{equation}
This implies that at any given time the following ratio $R'$ holds between newly
formed stars that fall in the BSS and RGB selection boxes:
\begin{equation}\label{eq:diff}
R'=\frac{\displaystyle\int_{M_1}^{M_2} \phi(m) \textrm{d}m}{\displaystyle\int_{M(t)}^{\infty} \phi(m) \textrm{d}m} =
\frac{M_1^{-1.35}-M_2^{-1.35}}{(t/t_{\odot})^{0.45}}.\vspace{0.1cm}
\end{equation}
For practical purposes, the power law of the IMF implies that the upper
  limit of integration at the denominator goes quickly toward a negligible
  contribution, be it a few tens of solar masses or infinity.
If one wishes instead to derive the ratio $R$ between all stars formed until
a given time that fall onto the BSS and RGB boxes, it suffices to integrate
over $t$, and to correct for the stars that have evolved off the selection
boxes. For the BSS box, this implies removing all stars  that have evolved off
the main sequence, i.e. with masses in the range $M(t)$ to $M_2$. For the RGB
box, one should correct for all stars that have left the giant branch. At a
given age, the RGB spans a mass range that is of order of a few percent of
the turnoff mass, $p \sim 1.1-1.03$.
\begin{displaymath}
R=\frac{\displaystyle\int_0^t \left[ \displaystyle\int_{M_1}^{M_{2}}\phi(m)\textrm{d}m - \displaystyle\int_{M(t)}^{M_{2}}\phi(m)\textrm{d}m \right]\,\textrm{d}t }{\displaystyle\int_0^t \displaystyle\int_{M(t)}^{p\,M(t)} \phi(m) \textrm{d}m\,\textrm{d}t}
\end{displaymath}
\begin{equation}\label{eq:int}
=\frac{1.45\,M_1^{-1.35}\,- (t/t_{\odot})^{0.45}}{(1-p^{-1.35})\,(t/t_{\odot})^{0.45}}.
\end{equation}
Here the integration over time corresponds to assuming a constant star-formation
history, which suffices to describe most of the evolution of the Galactic disk,
at least over the past $\sim 8$~Gyr \citep[e.g.,][]{snaith}
The dependence of equation \ref{eq:int} with time is shown in the left panel
of Figure \ref{fig:mcanal}, where I have used isochrones for an informed
guess on $M_1$. I have adopted the MIST isochrones \citep{choi} with
$-2.0\le \rm{[Fe/H]} \le 0.5$ \citep[roughly the range of metallicities
  covered by stars with $V_{\rm{TAN}}<200\kms$, e.g.,][]{sahl} and with ages
spanning over the entire grid of MIST isochrones. I have then identified all
isochrone masses that fall in the BSS box of Figure \ref{fig:video}, obtaining
the values $M_1=1.0$ and $M_2=1.8$ for the 10th and 90th percentiles,
respectively.

Stars with $V_{\rm{TAN}}>200\kms$ typically have ages older than $\sim 10$~Gyr
and
belong to the kinematically hot tail of the thick disk and to the stellar halo
\citep[e.g.][]{helmi,dimatteo,sahl}.
The simple model of equation \ref{eq:int} is sufficient to inform that at these
old ages there will not be residual main-sequence stars falling
into the BSS box. This reinforces the interpretation that the majority of stars
identified in the BSS box with $V_{\rm{TAN}}\gtrsim 200\kms$ are
genuine blue stragglers. 

The constant ratio of BSS at high $V_{\rm{TAN}}$ can also be qualitatively
understood. I assume $0.8 M_{\odot}$ as the typical stellar mass for halo
(sub)giant
stars \citep[e.g.][]{vdb,eps}, and I further assume that if a star occupies the
BSS box, it must roughly be $>1 M_{\odot}$ (as supported by the isochrones check
done above). The number of BSS with final mass $>1 M_{\odot}$ is given by a
fraction of all possible combinations of mass for stars between $M_a=0.2$ and
$M_b=0.8 M_{\odot}$, and which are in a binary system. Adopting the
\cite{kroupa01} IMF (whose broken power law is more appropriate than the
Salpeter one for masses below $M_K=0.5 M_{\odot}$) leads to:
\begin{displaymath}
R = f_c f_m \frac{\displaystyle\int_{t_2}^t \displaystyle\int_{M_a}^{M_b}\phi(m)\textrm{d}m\,\textrm{d}t}{\displaystyle\int_{t_1}^t \displaystyle\int_{M(t)}^{p\,M(t)} \phi(m)\textrm{d}m\,\textrm{d}t}=
\end{displaymath}
\begin{displaymath}
  1.43 f_c f_m t_{\odot}^{0.43} \frac{\left(t-t_2\right)}{\left(t^{1.43}-t_1^{1.43}\right)} \times
\end{displaymath}
\begin{equation}\label{eq:kroupa}
  \frac{\displaystyle\frac{4}{15}\left(M_a^{-0.3}-M_K^{-0.3}\right)+\displaystyle\frac{2}{65}\left(M_K^{-1.3}-M_b^{-1.3}\right)}{\displaystyle\frac{2}{65} (1-p^{-1.3})}.
\end{equation}

Here the number of RGB stars is given by the number of objects in the
appropriate mass range that formed at least $t_1=9$~Gyr ago, where this age
corresponds to a turnoff mass of $\sim 0.8 M_{\odot}$ (equation
\ref{eq:kippe}). Moving to the number of BSS, I do not make any assumption
when their mergers occur (which in fact could happen at times more recent than
$t_1$), nor their formation channels (e.g., binary interaction when both
components are still on the main sequence, but also from interaction with an
evolved primary with a mass somewhat higher than $M_b$). There is a degree of
stochasticity in the time at which each binary merger will occur, besides the
effect of stellar lifetime in the range $M_a$ to $M_b$, and of stragglers in
the range $M_1$ to $M_2$ (which will then evolve toward the giant phase).
Modeling these effects goes beyond the analytical formulation presented
here. The net effect of decreasing the number of BSS at any given time can
be expressed as a delay, where the number
of BSS will start to increase linearly from $t_2>t_1$. Choosing, e.g.,
$t_2=10$~Gyr produces a smooth slope similar to what is seen at $\vtan$
between $200$ and $300\kms$. For $t_2=t_1$ the rise is much steeper, although
irrelevant for the qualitative sake of this discussion.

Only a certain fraction of all possible mass combinations in the range
$0.2-0.8 M_{\odot}$ will result in a sum $>1 M_{\odot}$, and this is accounted 
for by $f_c$. This correction factor is determined numerically, by generating a
distribution of masses according to the adopted IMF, and for a given fraction
of binaries $f_b$ computing how many will have a total mass $>1 M_{\odot}$. This
returns $f_c \sim 0.18 f_b$, where $f_b$ can be taken from observations.
The  only free parameter is thus $f_m$, which is the fraction of binaries
undergoing
mass transfer/merger. This can be determined by requiring the plateau at old
ages of Equation \ref{eq:kroupa} to match that observed in the  halo
(Figure \ref{fig:mcanal}). Decent agreement is obtained if
$f_c f_m= 0.18 f_b f_m = 0.018$. Adopting $f_b=0.5$ for the fraction
of metal-poor close binaries \citep{mo} implies $f_m=0.2$, i.e. 20 percent
  of close binaries (or equivalently 10 percent of stars) will undergo some
sort of mass transfer and/or merger. Note that different values of binary
fraction will vary the percentage of binaries undergoing mass transfer
(e.g., $f_b=0.4$ implies $f_m=0.25$), but the total fraction of stars
($f_b f_m$) remains unchanged at 10 percent.

The purpose of this analytical formulation is simply to show that with a few
basic assumptions on the IMF and stellar lifetimes, it is possible to
qualitatively describe the trend seen in the BSS-to-RGB ratio of Figure
\ref{fig:video} and \ref{fig:mcanal}. At low $\vtan$ (i.e., young and
intermediate age stellar populations) the trend reflects the number of
main-sequence versus red giant branch stars. The flattening seen at high
$\vtan$ (old populations) can instead be described assuming that stars in the
BSS box are created by stellar mergers with a set of reasonable parameters.
\begin{figure*}
\epsscale{1.2}
\plotone{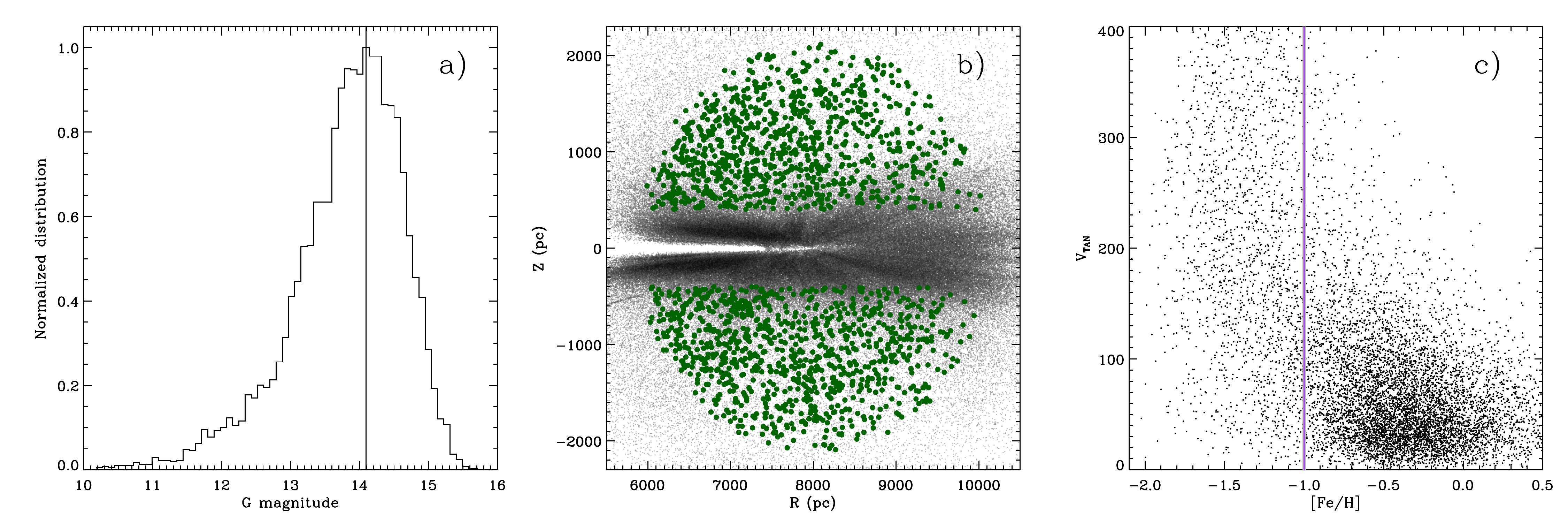}
\caption{{\it Left panel:} distribution of {\it Gaia} $G$ magnitudes for stars
  into the bRGB box, with $|Z|>400$~pc and $V_{\rm{TAN}}>250\kms$.
  The vertical line is the mode of the distribution. {\it Middle panel:} gray
  density plot is the distribution of all stars into the bRGB box as
  a function of height from the Galactic plane $Z$, and Galactocentric distance
  $R$ \citep[assuming that the Sun is a 8 kpc from the Galactic center,][]{gravity}.
  Green circles are bRGB stars with $|Z|>400$~pc and $V_{\rm{TAN}}>250\kms$ and closer
  than $\sim 2$~kpc. {\it Right panel:} [Fe/H] versus $\vtan$ for bRGB stars
  in the SkyMapper sample with $|Z|>400$~pc and closer than $\sim 2$~kpc. Stars
  with metallicities below $-1$ (vertical line) are classified as halo. 
\label{fig:3pan}}
\end{figure*}

\section{Building a volume complete sample of halo blue stragglers}
\label{sec:norma}

Membership to the stellar halo based only on $V_{\rm{TAN}}$ is rather
approximate. \cite{gaiaHR} have revealed that stars with $V_{\rm{TAN}}>200\kms$
fall along two well-defined sequences separated by roughly $0.1$ magnitude
in color, dubbed the red and blue sequence.
At $V_{\rm{TAN}}=200\kms$ there is still a clear contribution of thick disk
stars falling onto the red sequence, and the stellar halo becomes clearly
dominant only above $V_{\rm{TAN}}=250-300\kms$ \citep[see Figure 5 in][]{sahl}.
The blue sequence has been speculated to be formed by stars accreted by one
(or more) massive dwarf galaxy, whereas the red sequence likely comprises the
tail of the thick disk, kinematically heated by the accretion event
\citep[e.g.,][]{helmi,haywood,dimatteo,sahl,koppe,mye}.

The BSS-to-RGB ratio in the left panel of Figure \ref{fig:mcanal} bottoms
off and reverses between 200 and $300\kms$, after which it stabilizes to a mean
(and median) value of $0.20$ with a standard deviation of $0.03$. This value
is remarkably robust and well within the range measured by \cite{santana} in
dwarf galaxies with no recent star formation ($0.29$ with a standard deviation
of $0.17$), with selection boxes similar to those adopted here\footnote{As in
  Santana et al. (2013), I use the BSS-to-RGB ratio ($R$) at the face value.
  The actual value will be slightly higher, $R/(1-R)$ under the assumption that
  a fraction $R$ of RGB stars are in fact evolved BSS. At the same
  time, the measured $R$ is an upper limit, since the adopted box
  does not extend to the tip of the RGB.}. Whether or not the local halo is
  formed by one or more disrupted dwarf galaxies, it might not come as a
  surprise that the density of BSS in low metallicity, low density
  environments, such as the Galactic halo and dwarf galaxies, is similar
  \citep[see, e.g.,][]{momany07}.

\cite{santana} report a correlation between the number of BSS in a dwarf
galaxy and the total stellar mass of the system (their equation 5). Motivated
by the constant BSS-to-RGB ratio found at high tangential velocities and its
similarity to that measured in dwarf galaxies, I use the aforementioned
correlation to test whether it returns a sensible estimate for the stellar mass
in the local halo. To successfully doing so, it is crucial to correctly
assign BSS to the halo. This is not trivial to do purely based on $V_{\rm{TAN}}$
because of contamination from the thick disk above $200\kms$, and the fact
that the halo extends below this velocity. The blue sequence of the {\it Gaia}
H-R diagram offers a way out.

I define an RGB to be a member of the {\it Gaia} blue sequence (bRGB) if it
falls on the green box of Figure \ref{fig:video}. This box is contained within
the RGB box, and its boundaries on the right-hand side have been defined
selecting the midpoint where the Gaia red and blue sequences are most
separated in the H-R diagram of high $\vtan$ stars (in a fashion similar to
\citealt{sahl}. This is best appreciated in the animation of Figure
\ref{fig:video}). This region encompasses mostly old, metal-poor red giants
from the halo, although younger and more metal-rich red giant stars from the
Galactic disk can contaminate it, especially at low $\vtan$. I use
metallicities from the SkyMapper photometric survey \citep{c19} to identify
halo stars purely from their chemistry, and I estimate the fraction of missing
halo stars when cutting at a given $\vtan$ and height $Z$. With this
correction, I then derive a volume complete number of halo bRGB, and use the
constant BSS-to-RGB ratio of Figure \ref{fig:mcanal} to estimate the number of
halo blue stragglers, and total stellar mass within the same volume through
equation 5 of \cite{santana}. It can be seen from the animation of Figure
\ref{fig:video} that at the highest $\vtan$ (where the halo sample is the
cleanest, with all RGB stars virtually on the blue sequence) the BSS-to-RGB
ratio remains constant at $0.2$. This indicates that contamination from thick
disk stars affects equally BSS and RGB stars (not unexpectedly, since they
have similar intrinsic luminosities and hence probe similar distances), and
their ratio is thus a robust quantity.

Below, all these steps are explained in detail with a case study, using a set
of fixed parameters. This procedure is then generalized using 100,000
Monte Carlo realizations, where the parameters adopted in the case study are
changed within a reasonable range.
\begin{figure*}
\epsscale{1.2}
\plotone{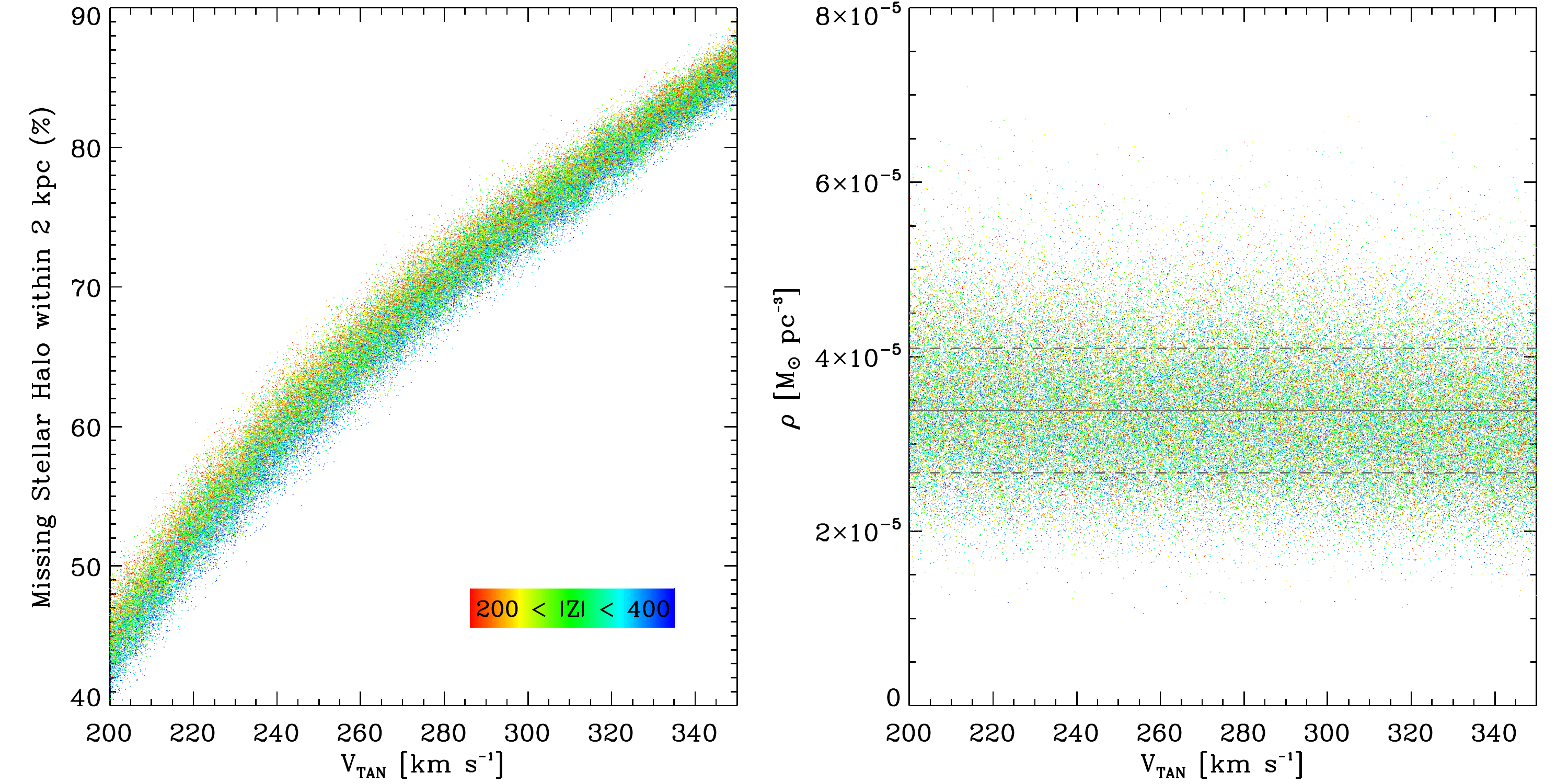}
\caption{{\it Left panel:} fraction of missing halo bRGB stars as a function of
  $\vtan$, for different cuts in height $Z$ above the Galactic plane (shown in
  color). Each
  point is one of the 100,000 Monte Carlo realizations of the procedure
  described in the text. {\it Right panel:} stellar halo density from each
  realization. Gray lines mark the mean (solid line) and one sigma levels
  (dotted lines).
\label{fig:rho}}
\end{figure*}

{\it Gaia} DR2 parallaxes have a typical precision of order $0.04$~mas for
stars brighter than $G\sim15$ (which is appropriate for the bulk of this
sample). This, together with the requirement of 10 percent precision in
parallaxes (Section \ref{sec:data}) limits completeness to parallaxes
$\gtrsim0.4$~mas (or distances closer than about $2.5$~kpc). This estimate,
however, does not account for the fact that stars with different intrinsic
luminosities will be complete to different distances. The distribution of $G$
magnitudes for bRGB shows the typical power law of a uniform, volume complete
sample up to $G\simeq14.1$ (Figure \ref{fig:3pan}a).
The green box of Figure \ref{fig:video} requires a complete sampling of bRGB
stars to be sensitive to $M_G\sim2.5$, thus implying that
$D=10^{\frac{14.1-2.5}{5}-2}\sim 2$~kpc is the farthest distance at which the
sample is complete. At bright magnitudes
{\it Gaia} DR2 is complete down to $G\sim7$ \citep[see discussion in][]
{gaiaDR2,bb}, which together with the bright limit of $M_G\sim0.6$ for the
bRGB box, translates to a distance completeness of $\sim 190$~pc. This limit
is
of no concern since I only select stars with $|Z|>400$~pc, to avoid regions
heavily affected by reddening and crowding, as well as strong contamination
from the disk (this cut in $Z$ eliminates all bRGB within $\sim 10^{\circ}$ from
the Galactic plane). There are 1824 bRGB satisfying these criteria.

SkyMapper provides [Fe/H] for some 9 million stars in the southern sky,
with no selection other than having good photometry, in a color range that
well encompasses the RGB box, and Galactic latitudes $|b|>5^{\circ}$ \citep{c19}.
All stars from the SkyMapper sample are in {\it Gaia} DR2: I apply the
quality flags and reddening corrections described in Section \ref{sec:data}
and identify SkyMapper members of the {\it Gaia} blue sequence with $|Z|>400$ pc
using the same green selection box of Figure \ref{fig:video}. Also for this
sample restricting to distances closer than $2$~kpc is appropriate (the
distribution of $G$ magnitudes for the SkyMapper sample peaks at a value
similar to that of the {\it Gaia} sample). Figure \ref{fig:3pan}c shows that
within the bRGB box there is a considerable fraction of metal
rich giants, as well as metal-poor stars with $\vtan<250\kms$. I classify a
star as halo if its [Fe/H]$<-1$, and I define the following correction for the
fraction of missing halo stars:
\begin{equation}
f_{ma}=1-\frac{n_{\vtan>250}^{|Z|>400}}{n_{[Fe/H]<-1}^{|Z|>400}},
\end{equation}
where the numerator and denominator are the number of bRGB stars 400~pc above
the Galactic plane with tangential velocities above $250\kms$ and metallicities
below $-1$, respectively. I find that $f_{ma}\sim 0.6$, i.e., within $2$~kpc
about 60 percent of halo stars are lost when cutting at $\vtan>250\kms$ and
$|Z|>400$pc. Thus, the complete number of halo bRGB is of order 4500.
  The choice of using [Fe/H]$<-1$ to chemically assign stars to the halo is
  arbitrary, and the transition from the thick disc and halo is not
  clear-cut \citep[e.g.,][]{rl08,ruchti10,sahl}. Nevertheless, from Figure
  \ref{fig:3pan}c it is clear that at the highest $\vtan$ (where the fraction
  of genuine halo stars is the highest) the bulk of stars has [Fe/H]$<-1$.
  Also, the adopted choice is consistent with the literature,
  where the broad metallicity distribution of the halo is found to become
  prominent below $-1$ \citep[e.g.,][]{rn91,an13}, whereas at high $\vtan$ the
  thick disk peaks at $-0.7$ \citep[][]{sahl}.
  
It must be noted that the completeness of the samples is partly decreased by the
quality cuts described in Section \ref{sec:data}. To assess their effect, I
query the {\it Gaia} archive requiring only parallaxes better than 10 percent.
This results in a sample of about $72.5$ million objects, i.e. about 7 percent
larger than the one used in Section \ref{sec:data}. This is consistent
with the order 10 percent effect found by \cite{bb} when introducing quality
cuts on a sample with parallaxes better than 20 percent. If I only consider
stars in a color range broadly consistent with the location of the of BSS
($0<G_{{\rm BP}}-G_{{\rm RP}}<0.8$) and RGB ($0.7<G_{{\rm BP}}-G_{{\rm RP}}<1.3$)
boxes, the mean and median difference
of the two samples as a function of Galactic latitude is 6 percent, with a
scatter of 4 percent. This check is to ensure the absence of significant
trends with latitude, due to the fact that the quality flag
{\tt phot\_bp\_rp\_excess\_factor} is sensitive to increasing stellar crowding
toward the plane of the Galaxy \citep{Gaia_evans}.

I thus increase by 6 percent the number of halo bRGB, and convert those into
the expected number of halo BSS using a fraction of 0.20 from Figure
\ref{fig:mcanal}. Using equation 5 from \cite{santana}, I estimate a total
halo stellar mass of $0.9\times10^6M_{\odot}$ within $\sim2$~kpc from the Sun.
Accounting for the volume of a missing spherical segment of height $\pm Z$:
\begin{equation}
V = \frac{\pi Z}{3}(6D^2-2Z^2),
\end{equation}
where $D=2$~kpc and $Z=0.4$~kpc, returns a local stellar halo density
$\rho= 3.4 \times 10^{-5} M_{\odot}/\rm{pc}^3$.

The procedure outlined above is repeated 100,000 times, varying each time
with Gaussian random errors the boundaries of the bRGB box (Figure
\ref{fig:video}), reddening by 20
percent, parallaxes within their quoted errors, completeness correction by
$6\pm4$ percent, and imposing different cuts in $200<|Z{\rm(pc)}|<400$ and
$200<\vtan(\kms)<350$. A correction for the fraction of missing
halo bRGB is determined each time, by similarly perturbing for the SkyMapper
sample reddening, parallaxes, and metallicities by $0.2$~dex. The mode of the
distribution of $G$ magnitudes is determined for the {\it Gaia} and the
SkyMapper sample, and the brightest of the two is used to find the farthest
distance at which both samples are complete given the faintest $M_G$ of the bRGB
box used. The correction for the fraction of missing halo bRGB is applied to
derive the actual number of halo bRGB, which is then converted into a number
of BSS using a ratio of $0.20\pm0.03$.

From the procedure described above, I obtain the following value for the
average stellar halo density within $2$ kpc from the Sun (Figure \ref{fig:rho}):
\begin{equation}
\rho=(3.4 \pm 0.7) \times 10^{-5} M_{\odot}/\rm{pc}^3.
\end{equation}
While the correction for the fraction of missing halo bRGB stars is a
strong function of $\vtan$, $\rho$ is remarkably flat, as one would expect
after applying a proper completeness correction. A wide range of density
normalizations have been found in the literature \citep[$3-15\,\times 10^{-5} M_{\odot}/\rm{pc}^3$][]{morrison93,fj98,gould98,dj10}. The value derived here
compares more favorably to low normalizations, although the very different
values obtained by different authors over the years highlight how difficult it
is to derive a definitive measurement. 

\begin{figure}
\epsscale{1.2}
\plotone{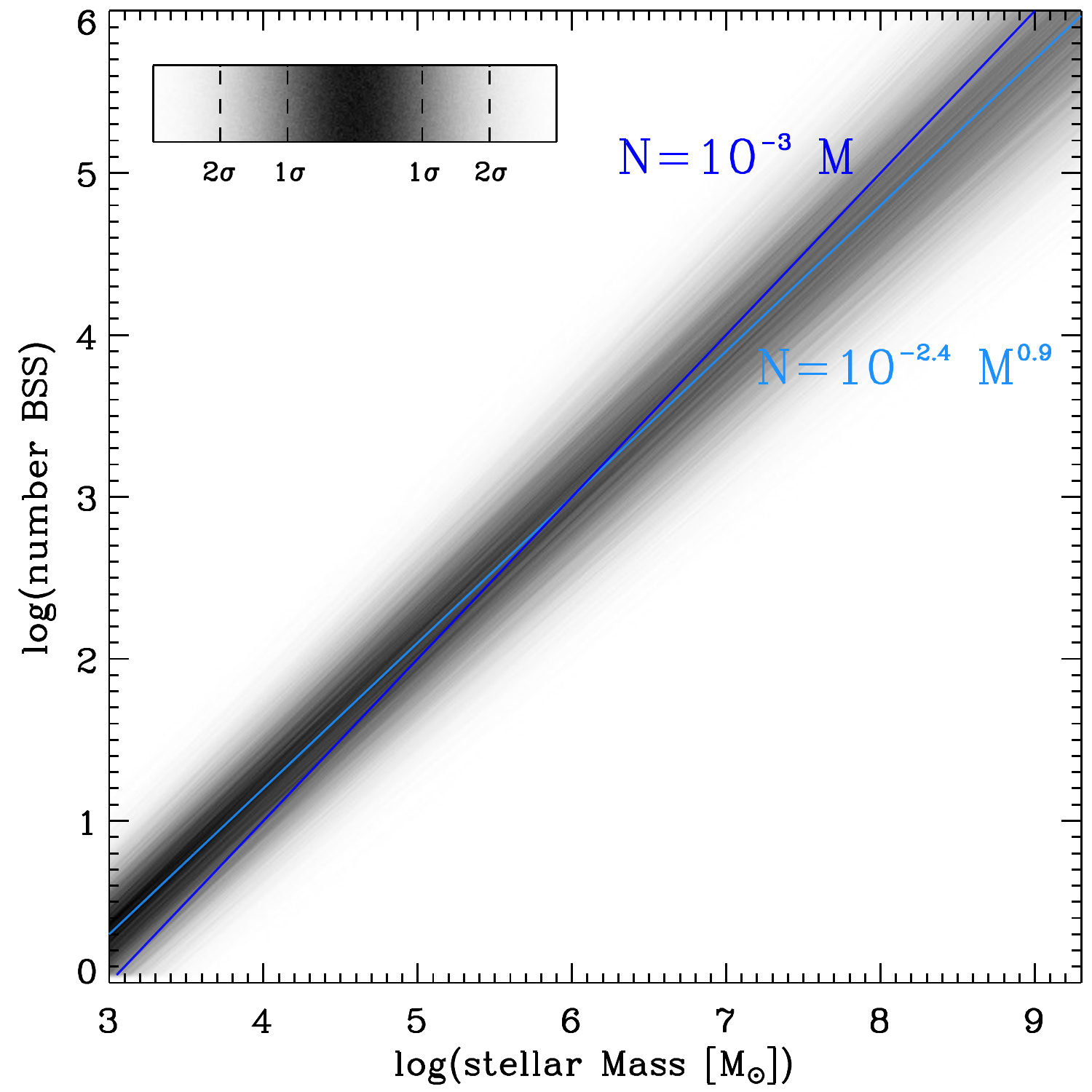}
\caption{Relation between the number of BSS and the stellar mass of a dwarf
  galaxy from \cite{santana} (light blue). The region allowed by the
  uncertainties of the relation is shown in gray (with 1 and 2
  $\sigma$ levels indicated in the upper bar). A linear relation that still
  fits within the errors is shown by the dark-blue line. 
\label{fig:fit}}
\end{figure}

It must be noted that the relation of \cite{santana} is not exactly linear
between the number of BSS and stellar mass. In other words, the stellar
mass of a dwarf galaxy with $n$ blue stragglers is different from the mass of
$m$ dwarf galaxies, each containing $i$ blue stragglers $\sum i=n$. In the
extreme (and unrealistic) case that each blue staggler comes from a different
dwarf galaxy, the difference with respect to assuming all from the same dwarf
amounts to $1-n^{-0.11}$, or about 50 percent for $n \sim 10^3$. 
Recent evidence suggests that the local Galactic halo is the result of two or
few massive mergers \citep[e.g.,][]{mye}, in which case the difference reduces
to the order of some percent. However, we do not know the total number of blue
stragglers, of which only a fraction is observed within 2 kpc. Fortunately,
Figure \ref{fig:fit} shows that the relation of \cite{santana} can be well
approximated with a linear function within its uncertainties. Adopting the
linear form of Figure
\ref{fig:fit}, the inferred stellar halo density increases at the percent
level only. It should be pointed out that the relation of \cite{santana} is
calibrated between $\sim10^3$ and $\sim10^6 M_{\odot}$, and here it has been
applied within this range. The comparison of the linear function in Figure 
\ref{fig:fit} extends up to $10^9 M_{\odot}$ \cite[the total stellar mass of the
  halo; e.g.,][]{deason}. While it must be explored whether the adopted linear
relation holds to this regime, the point here is that while different linear
functions will change somewhat the stellar mass derived, the effect is within
the quoted uncertainties. 

\section{Scaling relations}\label{sec:bho}

\begin{figure*}
\epsscale{1.2}
\plotone{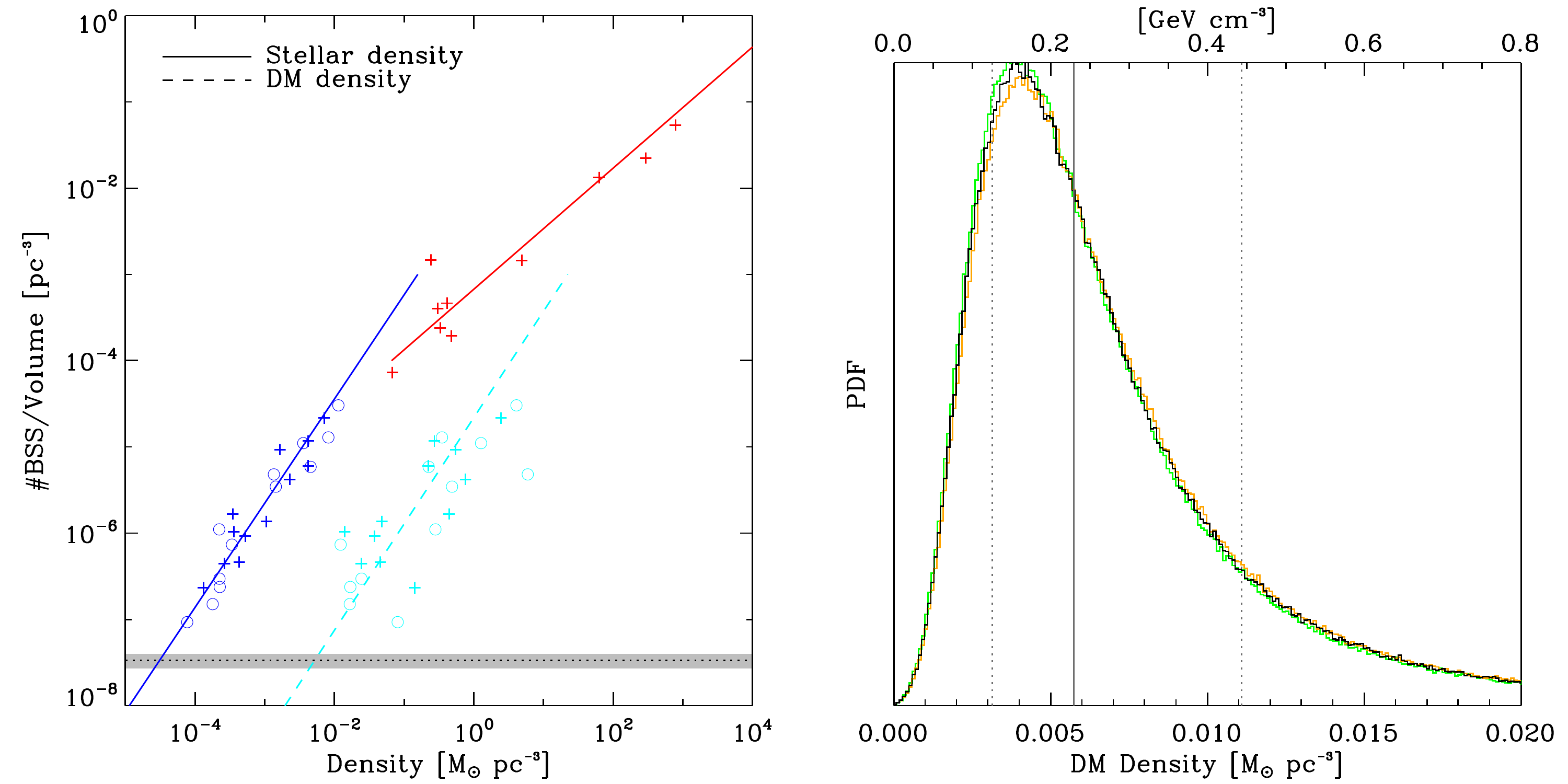}
\caption{{\it Left panel:} volume density of BSS as a function of stellar density
  in dwarf galaxies (blue) and globular clusters (red). The volume density of
  BSS is found to correlate also with the dark matter density in dwarf
  galaxies (cyan). Plus signs (open circles) have been derived using data from
  \cite{munoz18} \citep{mcc12} as described in the text. All densities are
  computed within a half-light radius. The dotted line is the volume density of
  halo BSS within 2 kpc
  from the Sun (with $1\sigma$ error from the Monte Carlo described in Section
  \ref{sec:norma}). {\it Right panel:} probability distribution function for
  the dark matter density within $2$~kpc of the solar location. Gray vertical
  lines are the median (solid), 16 and 84 percent values (dotted).
  Lines of different colors assume different stellar mass-to-light ratios (1
  orange; $1.5$ black; 2 green).
\label{fig:scaling}}
\end{figure*}

The good agreement of the local baryon density with estimates from the
literature from the previous section warrants further investigation of whether
scaling relations derived from BSS in dwarf galaxies can be
applied to the Milky Way halo. Figure \ref{fig:scaling} shows the volume
density of blue stragglers as a function of stellar density for both dwarf
galaxies (solid blue line) and globular clusters (solid red line) from
\cite{santana}. Stellar densities have been calculated within a half-light
radius, using $0.5\,L_{V}/L_{\odot,V}$ \citep[from][]{munoz18}, and
assuming a
stellar mass-to-light ratio of $M_{\star}/L_{V}=1.5 M_{\odot}/L_{\odot,V}$. This
value is appropriate for both dwarf galaxies and globular clusters in this
sample \cite[e.g.,][]{woo08,bau}. I have used half-light radii from S\'ersic
profiles $R_{h,s}$ since in \cite{munoz18} those are available for both
dwarf galaxies and globular
clusters, and note here that differences are negligible for systems having
half-light radii from exponential or Plummer profiles. I multiply by $2/3$ the
number of BSS in \cite{santana} since those are counted up to two half-light
radii\footnote{Assuming for simplicity an
  exponential profile $e^{-kx}$ where $k$ is an integer, there is $3/4$ of
  light within two half-light radii, and $2/3$ of this $3/4$ is within a
  half-light.}. 

The density of BSS steeply correlates with stellar density in both dwarf
galaxies and globular clusters. For dwarf galaxies, the intercept
of the relation with the density of halo BSS determined in
the previous section returns a stellar density of
$(3.1 \pm 0.5) \times 10^{-5} M_{\odot}/\rm{pc}^3$. The good agreement with the
value previously determined is not unexpected, both ultimately
depending on the same set of data. However, it must be noted that the choice of
comparing the
density of halo BSS to that within a half-light radius of dwarf galaxies is
arbitrary. If stellar densities were to be computed within two
half-light radii, the stellar density inferred for the halo would change to
$(2.6 \pm 0.6) \times 10^{-5} M_{\odot}/\rm{pc}^3$. Not unexpectedly, the
largest source of systematic uncertainty is the adopted stellar mass-to-light
ratio, where a change of $\pm 0.5 M_{\odot}/L_{\odot,V}$ affects stellar
densities by $\pm 1.0 \times 10^{-5} M_{\odot}/\rm{pc}^3$.

More interestingly, the density of BSS is found to correlate with the
dark matter density of dwarf galaxies, here computed again within a half-light
radius. This can be understood from the dark matter scaling laws in late-type
and dwarf spheroidal galaxies, where the dark matter halo core density
correlates with the absolute magnitude of a galaxy, i.e.,~roughly stellar mass
content \citep[][see also Figure \ref{fig:smd}]{kf16}. Here stellar mass
content is traced by BSS.
The dark matter density has been estimated using dynamical
mass-to-light ratios $M/L_{V}$ reported in \cite{munoz18}, from which 
$M_{\rm{DM}}=(M_{\rm{dyn}}-M_{\star})=(M/L_{V}-M_{\star}/L_{V})L_{V}/2$, where the
factor of 2 follows from the choice of working at half-light radius (i.e.
in the following $M_{\rm{DM}}$, $M_{\rm{dyn}}$ and $M_{\star}$ are all computed
within half-light radius).
The contraction of the dark matter halo due to the addition of stars
can be corrected by adiabatically expanding the half-light radius. Assuming
circular orbits and angular momentum conservation, this gives
\begin{equation}
R_{h,s}^{'}=\frac{M_{\star}+M_{\rm{DM}}}{M_{\rm{DM}}} R_{h,s}
\end{equation}
\citep[e.g.,][]{blu,for}, which has virtually no effect since for this sample of
dwarf galaxies $M_{\star}$ is orders of magnitude smaller than $M_{\rm{DM}}$.
The approach used here averages the dark matter density over a half-light
radius, which is appropriate if (dwarf) galaxies --as it seems-- have cored
profiles \citep[e.g.,][]{serra10,read16,li20}.
The derived $M_{\rm{DM}}$ does not account for the effect of tidal stripping
of the halos of dwarf galaxies.  Nevertheless, this simple methodology
returns dark matter densities that typically agree to within a few tens of
percent with the values derived from the detailed modeling of
\cite{read19} for the same galaxies in their sample. 

To account for uncertainties, I have repeated the above procedure to derive
stellar and dark matter densities using half-light radii, stellar and
dynamical masses from \cite{mcc12}. Differences with respect to the values
obtained using mass-to-light ratios from \cite{munoz18} are typically of few
tens of percent. I then generate a million realizations
building each time a sample that randomly mixes data from \cite{mcc12} and
\cite{munoz18}, and I perturb them by a Gaussian of width equal to half of their
differences. The intercept with the local density of halo BSS (also perturb
within its uncertainties) returns a median dark matter density:
\begin{displaymath}
  \rho_{DM}= 0.0058^{+0.0053}_{-0.0026}\, M_{\odot}/\rm{pc}^3  
\end{displaymath}
\begin{equation}
  =0.22^{+0.20}_{-0.10}\,\rm{GeV}/\rm{cm}^3.
\end{equation}
The mean density is instead $0.0076 M_{\odot}/\rm{pc}^3 =
0.29\,\rm{GeV}/\rm{cm}^3$. In can also be appreciated from Figure
\ref{fig:scaling} that varying the stellar mass-to-light ratio by
$\pm 0.5 M_{\odot}/L_{\odot,V}$ has a negligible impact on the inferred dark
matter density. 

The median (mean) value of $\rho_{DM}$ determined here is in overall good
agreement with those recently reported in the literature, which are in the range
$0.005-0.013\,M_{\odot}/\rm{pc}^3$ \citep[$\simeq 0.2-0.5\,\rm{GeV}/\rm{cm}^3$,
  see, e.g.,][]{swe12,z13,bt12,br13,mckee15,pmm17,s18,ds19}. These methods are
based on dynamically modeling the rotation curve or the vertical motion of
stars, and in either case a number of assumptions are needed. One of the most
important ones is
the contribution of baryonic matter to the local dynamical mass, which is
nontrivial to determine and strongly correlates with the inferred dark matter
density \citep[e.g.,][]{flynn06,s18}. 

The use of BSS proposed here is largely independent of the baryonic
content (Figure \ref{fig:scaling}, right panel), and the mass estimators used
to infer dynamical masses within the half-light radius of dwarf galaxies are
believed to be accurate \citep{camp17,gosa}. The choice of applying a scaling
relation inferred from dwarf galaxies to estimate the local dark matter halo is
motivated by the similar BSS-to-RGB ratio measured in the local halo compared to
dwarf galaxies. Admittedly, however, in dwarf galaxies these scaling relations
are estimated at half-light radii, whereas here they are applied to field
stars as a whole. 

\section{Discussion and Conclusions}\label{sec:end}

The correlation between stellar mass density and volume density of BSS shown in
Figure \ref{fig:scaling} can be readily understood from the findings of
\cite{santana} i.e., the number count of BSS increases with the stellar mass
content of dwarf galaxies, whereas it stays constant in globular clusters. With
increasing stellar mass, globular clusters are typically more compact. This
means that the volume density of BSS increases with stellar density, moving
from the bottom left to the top right of the red line. For dwarf galaxies,
the pathway is opposite. With increasing stellar mass content, the number of
BSS in dwarf galaxies increases, and so do half-light radii. This leads to a
decrease of both stellar mass density and BSS volume density with increasing
galaxy mass,
i.e. moving from the top right to the bottom left of the blue curve. Because
the stellar mass content of dwarf galaxies correlates positively with their
dark matter content, the same trend still holds when dark matter density is used
instead of stellar density. Whether the correlation between the density of BSS
and that of dark matter is indicative of a connection between baryons and
dynamics is something worth contemplating \citep[e.g.,][]{sancisi,stacy}, but
beyond the scope of this paper.

\begin{figure}
\epsscale{1.2}
\plotone{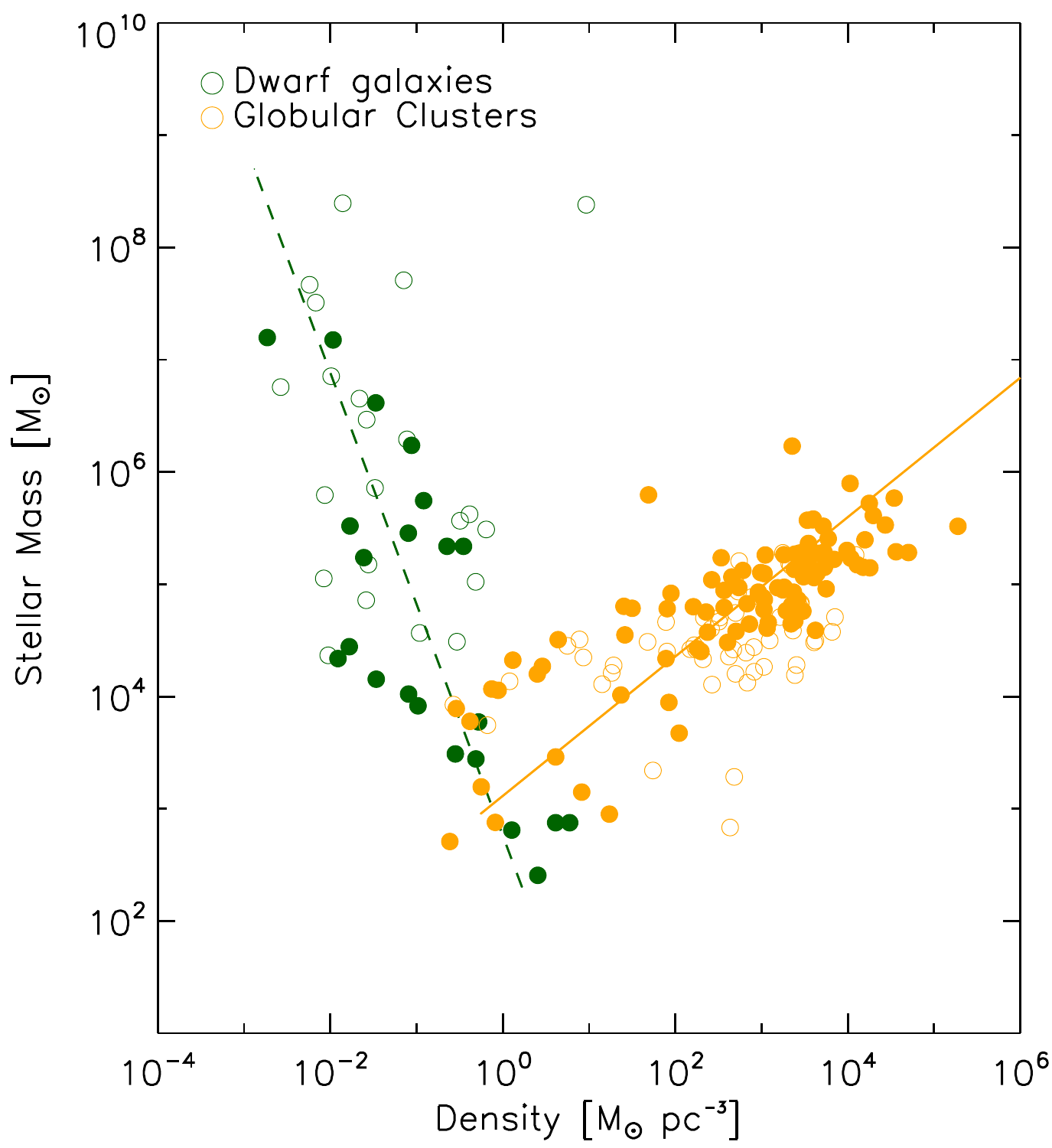}
\caption{Stellar mass vs. density relation, using the compilation of dwarf
  galaxies from \cite{mcc12} and globular clusters from \cite{bau}. For dwarf
  galaxies, the dark matter density is plotted, computed as described in
  Section \ref{sec:bho}. For globular clusters, the stellar density is plotted.
  Densities and stellar masses are both computed within a half-light radius.
  Filled circles are dwarf galaxies of the Milky Way (green) and globular
  clusters (yellow) with masses better than 20 percent.
  \label{fig:smd}}
\end{figure}

The constant number of BSS as a function of cluster mass ($M_{\star}$) translates
into a decreasing number of BSS per unit mass
($\propto 1/M_{\star}$), whereas the number of BSS per unit mass stays roughly
constant in dwarf galaxies ($\propto M_{\star}^{\alpha} \times 1/M_{\star} \sim
\rm{const}$), where $\alpha \sim 1$ (see Figure \ref{fig:fit}). The number of 
BSS per unit mass can be interpreted either as a measure of formation
or disruption efficiency of BSS. For example, if BSS are the product of close
binaries, it could be argued that in denser stellar systems close binaries are
less likely to form, or that closer binaries are more easily disrupted.
While addressing these questions is beyond the scope of this paper
\citep[see, e.g.,][for a review]{momany15}, I note that
the trends discussed here using BSS can be traced in the stellar mass versus
density relation (Figure \ref{fig:smd}). This relation is equivalent to the
more popular absolute magnitude vs. half-light radius relation, and shows how
the stellar mass of a system has a positive correlation with stellar density
in globular clusters, and a negative correlation with dark matter
density in dwarf galaxies. BSS are thus tracing these scaling relations, and
here I have applied them to the Milky Way halo. 

Further investigations are needed to confirm the use BSS as a proxy of baryon
and dark matter density in the halo, as this technique could be very
powerful, e.g., at measuring these quantities across the Milky
Way halo on the same scale as in external galaxies. Remarkably, applying to the
local Milky Way halo scaling relations inferred for BSS in dwarf galaxies is
able to return both a stellar density and dark matter density that are in
overall good agreement with other
determinations in the literature. As discussed in the paper, BSS thus seem to
trace stellar mass in low density, low metallicity environments regardless if
in dwarf galaxies or in the halo. When it comes to dark matter, the correlation
between the dark matter core density and stellar mass seen in dwarf galaxies
returns a meaningful result also for the local halo. It thus seems that the
nearby halo of a bright spiral like the Milky Way can lie on some of the
scaling laws for dwarf galaxies. Whether this holds universally, or because the
nearby halo is largely formed by disrupted dwarf galaxies remains to be
seen.

\acknowledgments
I thank Rosemary Wise for an inspiring colloquium on blue stragglers that set
this work in motion, and Ken Freeman, Chris Flynn, Helmut Jerjen, Thomas
Nordlander, and Aldo Serenelli for a reading of the manuscript and comments.
Enrico di Teodoro, Ashley Ruiter, and Ivo Seitenzahl are acknowledged for
helpful discussions, and Felipe Santana for useful correspondence. I thank an
anonymous referee for constructive criticism, which has improved the paper.

Funding for this work has been provided by the ARC Future Fellowship
FT160100402. Parts of this research were conducted by the ARC Centre of
Excellence ASTRO 3D, through project number CE170100013.
This work has made use of data from the European Space Agency (ESA) mission
{\it Gaia} (\url{https://www.cosmos.esa.int/gaia}), processed by the {\it Gaia}
Data Processing and Analysis Consortium (DPAC,
\url{https://www.cosmos.esa.int/web/gaia/dpac/consortium}). Funding for the DPAC
has been provided by national institutions, in particular the institutions
participating in the {\it Gaia} Multilateral Agreement.

\vspace{5mm}
\facilities{Gaia, SkyMapper}

\bibliography{luca_v2}{}
\bibliographystyle{aasjournal}

\end{document}